\documentclass[intlimits,twoside,a4paper]{article}

\usepackage{graphicx}

\usepackage[cp1251]{inputenc}

\usepackage[eqsecnum]{cmpj3}

\DeclareMathOperator{\Tr}{Tr}

\issue{2017}{20}{2}{23301}
\doinumber{10.5488/CMP.20.23301}
\title[Investigation on nickel ferrite nanowire]%
{Investigation on nickel ferrite nanowire device exhibiting negative differential resistance –- a first-principles investigation }%
\author[V. Nagarajan, R. Chandiramouli]{V. Nagarajan, R. Chandiramouli\thanks{Corresponding author} }
\address{
School of Electrical and Electronics Engineering, Shanmugha Arts Science Technology and Research Academy (SASTRA) University, Tirumalaisamudram, Thanjavur, Tamil nadu --- 613 401, India}

\date{Received September 23, 2016, in final form December 13, 2016}

\begin{document}

\maketitle

\begin{abstract}
The electronic property of NiFe$_2$O$_4$ nanowire device is investigated through nonequilibrium Green’s functions (NEGF) in combination with density functional theory (DFT). The electronic transport properties of NiFe$_2$O$_4$ nanowire are studied in terms of density of states, transmission spectrum and $I$--$V$ characteristics. The density of states gets modified with the applied bias voltage across NiFe$_2$O$_4$ nanowire device, the density of charge is observed both in the valence band and in the conduction band on increasing the bias voltage. The transmission spectrum of NiFe$_2$O$_4$ nanowire device gives the insights on the transition of electrons at different energy intervals. The findings of the present work suggest that NiFe$_2$O$_4$ nanowire device can be used as negative differential resistance (NDR) device and its NDR property can be tuned with the bias voltage, which may be used in microwave device, memory devices and in fast switching devices.
\pacs 31.10.+z, 31.25.-v, 61.46.+w, 61.66.Fn, 73.63.Rt, 85.30.-z
\end{abstract}

\section{Introduction}
The spinel ferrite is one type of soft magnetic materials with the general formula of MFe$_2$O$_4$, where ``M'' represents the divalent metal ions such as Mg, Zn, Mn, Cu, Co, Ni, etc., which are the most attractive magnetic material owing to their significant magnetic, magnetoresistive and magneto-optical properties. The other fascinating characteristics of MFe$_2$O$_4$ are its low melting point, large expansion coefficient, low magnetic transition temperature and low saturation magnetic moment \cite{1}. In spite of these properties, the spinel ferrites have been utilized in many technical applications, such as in catalysis \cite{2}, photoelectric devices \cite{3}, nano-device \cite{4}, sensors \cite{5}, magnetic pigments \cite{6} and microwave devices \cite{7}. The remarkable magnetic and electronic property of ferrites mainly depends upon the cations, their charges and the distribution of cations along tetrahedral (A) and octahedral (B) sites \cite{8}. Nickel ferrite (NiFe$_2$O$_4$) is one of the most versatile materials due to its soft magnetic property, low eddy current loss, low conductivity, catalytic behaviour, high electrochemical stability, abundance in nature, etc., \cite{7}. NiFe$_2$O$_4$ is a kind of ferromagnetic oxide with inverse spinel structure in which Fe$^{3+}$ ions are equally distributed between both octahedral B-sites and tetrahedral A-sites, whereas Ni$^{2+}$ ions  occupy only octahedral B-sites \cite{9}. The inverse spinel ferrites are represented by the general formula of (Fe$^{3+}$)$_\text A$(Ni$^{2+}$Fe$^{3+}$)$_\text B$O$_4^{2-}$ \cite{10}. NiFe$_2$O$_4$ powders have been used as catalysts \cite{11}, ferrofluids \cite{12}, biomedicine \cite{13} and gas sensors \cite{14,15}. Various methods have been employed for the synthesis of nanoscale NiFe$_2$O$_4$, which includes solid-state reaction \cite{16}, sol-gel \cite{17}, rheological phase reaction method \cite{18}, mechanochemical \cite{19}, pulsed wire discharge \cite{20}, electrospinning \cite{21}, hydrothermal \cite{22} and sonochemical methods \cite{23}.

The nanoscale devices have attracted researchers and these devices may have high packing density and are more efficient than microelectronic devices. Moreover, the junction properties of nanoscale devices play a vital role in the charge transport across the semiconductor/metal interfaces \cite{24}. Furthermore, the semiconductor/metal interface may also form Schottky or ohmic contact. If Schottky type of contact is present, rectifying action takes place.  The transport characteristics of nanoscale contacts must be investigated before the amalgamation of these structures in nanoscale electronic devices \cite{25}. Transport properties of these nanoscale device contacts are also influenced by the charge carriers and the geometry of the semiconductor/metal interface. Negative differential resistance (NDR) behaviour is a most significant electronic transport property for various electronic components \cite{26}. The NDR effect can be observed from low dimensional nanostructures like nanowire when connected between two electrodes \cite{27}. In a negative differential resistance device, the occupied states on one side may get aligned with the gap on the other side, when the voltage across the device is increased. Moreover, the current reduction may also occur due to the position of the resonant states of the molecule, which move within the gap of one of the contacts. In the case of carbon nanotube junctions, the reduction in the current for an increased bias voltage is due to the mismatch in the symmetry of incoming and outgoing wave functions of the same energy. Besides, the NDR effect observed between gold electrodes and scattering region is due to the lack of orbital matching between the contacts. The potential barriers in 2D graphene sheets are due to the linear dispersion of electrons, which shows a gap in their transmission across the barrier \cite{28}. Thus, negative resistance provides a physical significance in nonlinear electronic components. NDR has attracted scientific community due to its vast applications in electronics, such as in oscillators, memory devices and fast switching devices \cite{29}. Nowadays, NDR has been demonstrated in various semiconductor systems, including molecular nanowire junctions \cite{30}, organic semiconductor \cite{31} and single electron devices \cite{32}. The NDR effect is associated with a variety of phenomena, including Coulomb blockade \cite{33}, tunnelling and charge storage \cite{34}. Ling \cite{35} reported the negative resistance property in triangular graphene p–n junctions induced by vertex B–N mixture doping. Liu and An \cite{36} investigated the negative resistance property in metal/polythiophene/metal structure. Chen \cite{37} investigated NDR in oxide-based resistance-switching devices. Gupta and Jaiswal \cite{38} reported NDR in nitrogen terminated doped zigzag graphene nano-ribbon field effect transistor. Zhao et al. \cite{39} studied NDR property and electronic transport properties of a gated C60 dimer molecule sandwiched between two gold electrodes. The inspiration behind the present work is to study the transport property of NiFe$_2$O$_4$ nanowire and to investigate its NDR property. In the present work, the transport characteristics of NiFe$_2$O$_4$ nanowire device and its NDR properties are explored at an atomistic level and the results are reported.

\section{Computational methods}
The first-principles calculation on inverse spinel NiFe$_2$O$_4$ molecular device is investigated through nonequilibrium Green’s functions (NEGF) in combination with density functional theory (DFT) method utilizing TranSIESTA module in SIESTA package \cite{40}. NiFe$_2$O$_4$ nanowire is optimized by reducing the atomic forces on the atoms in nickel ferrites to be less than 0.05~eV/{\AA}. The Brillouin zones of NiFe$_2$O$_4$ are sampled with $1\times1\times5$~$k$-points. The generalized approximation (GGA) along with Perdew-Burke-Ernzerhof (PBE) exchange correlation functional is used to study the electron-electron interaction \cite{41,42}. The negative differential resistance property of NiFe$_2$O$_4$ is also studied through SIESTA package, in which the core electrons are suitably replaced by Troullier-Martins pseudopotentials for nickel, iron and oxygen atoms. Moreover, the electronic wave functions of nickel, iron and oxygen atoms are demonstrated in terms of a basis set, which are mainly related to the numerical orbitals. The optimization of band structure and electronic properties of NiFe$_2$O$_4$ nanowire are implemented using the double zeta polarization (DZP) basis set for the right-hand, left-hand electrodes and the scattering region in the present study \cite{43}. In order to investigate the electronic properties of NiFe$_2$O$_4$ and to exclude the interaction of NiFe$_2$O$_4$ nanowire with its periodic images, 10~{\AA} vacuum padding is modelled along $x$ and $y$ axes. This makes the computation process easy while examining the density matrix Hamiltonian. The atoms in NiFe$_2$O$_4$ nanowire freely move along their respective positions until the convergence force smaller than 0.05~eV/{\AA} is achieved.
	
Sen et al. \cite{44} studied the transport properties of trimer unit of cis-polyacetylene and fused furan trimer using DFT in combination with NEGF \textit{ab initio} method. They observed the NDR over a bias voltage of ($+$2.1 to $+$2.45~V). Yu et al. \cite{45} investigated the transport properties of a few nm long single-walled carbon nanotube (SWCNT) p--n junctions using the \textit{ab initio} quantum method. The finding reveals that nm long SWCNT shows negative differential resistance. Song et~al. \cite{46} reported NDR behaviour in (8,0) carbon/boron nitride nanotube heterojunction. They report that under positive and negative bias, the variation in the localization of corresponding molecular orbital under the applied bias voltage leads to NDR behaviour. Mahmoud and Lugli \cite{47} studied molecular devices with negative differential resistance. The molecular device is composed of diphenyl-dimethyl connected to the carbon chain linked to gold electrodes. They observed NDR behaviour only for an odd number of carbon atoms in the chain between the gold electrodes. In the present work, NDR behaviour is observed along NiFe$_2$O$_4$ nanowire. The adopted method in the present work resembles the method used in the above mentioned literature, which confirms the reliability of first-principles study on NiFe$_2$O$_4$ nanowire molecular device. The novel aspect of the present work is NDR properties of NiFe$_2$O$_4$ nanowire device which is discussed in terms of density of states spectrum, transmission and $I$--$V$ characteristics.

\section{Results and discussion}
\subsection{Structure of NiFe$_2$O$_4$ nanowire}
\begin{figure}[!b]
\vspace{-4mm}
\centerline{\includegraphics[width=0.85\textwidth]{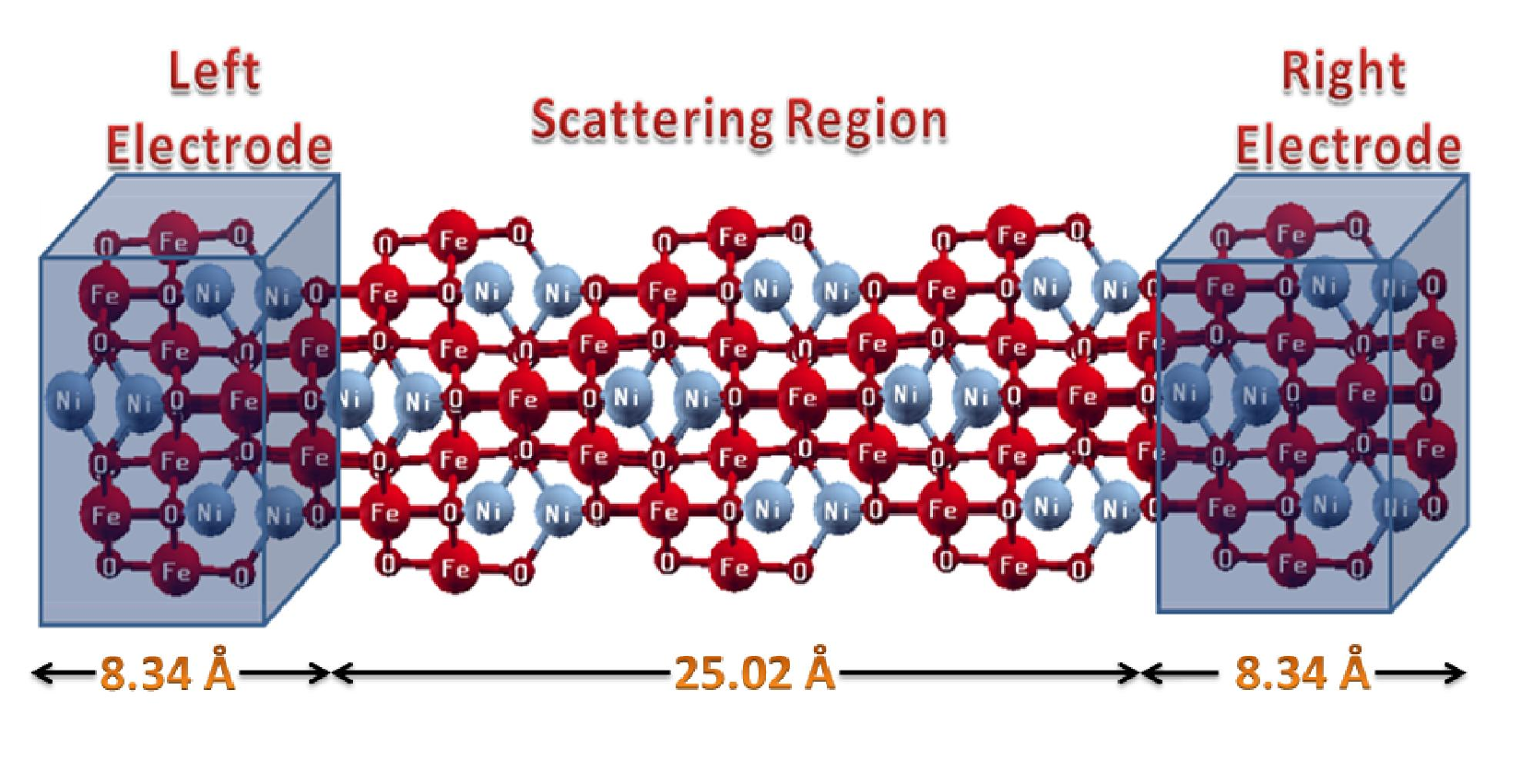}}
\caption{(Color online) NiFe$_2$O$_4$ molecular device.} \label{fig-s1}
\end{figure}

The NiFe$_2$O$_4$ nanowire device is built using International Centre for Diffraction Data (ICDD) Card number: 03-0875, which exhibits the inverse spinel structure. The designed NiFe$_2$O$_4$ nanowire molecular device is divided into three regions, namely left-hand electrode, scattering region and right-hand electrode regions. The scattering region of NiFe$_2$O$_4$ nanowire device is placed in between two electrodes. The corresponding width of the scattering region, left-hand electrode and right-hand electrode are 25.02~{\AA}, 8.34~{\AA} and 8.34~{\AA}. The NiFe$_2$O$_4$ nanowire is repeated five times along $c$-axis. Initially, in order to optimize the dimension of the molecular device, the NiFe$_2$O$_4$ molecular device is built with different dimensions. Moreover, when the dimension of the scattering region is small, it gives rise to the tunnelling of electrons across the NiFe$_2$O$_4$ device. However, if the dimension is too long, the magnitude of the current flowing across the NiFe$_2$O$_4$ device decreases. When NiFe$_2$O$_4$ device is of the order of the above mentioned dimensions, a significant current flows across the NiFe$_2$O$_4$ device.
Along NiFe$_2$O$_4$ scattering region, a bias voltage is maintained between the left-hand electrode and right-hand electrode for the flow of current. The scattering region of NiFe$_2$O$_4$ nanowire consists of twenty four nickel atoms, forty eight iron atoms and ninety six oxygen atoms.  The region on the left-hand and right-hand electrodes includes eight nickel atoms, sixteen iron atoms and thirty two oxygen atoms each. The potential difference of $-V/2$ and $+V/2$ is maintained across the right-hand and left-hand electrode in NiFe$_2$O$_4$ molecular device. Besides, the variation in the bias voltage leads to the change in the density of states and transmission along NiFe$_2$O$_4$ nanowire device. Figure~\ref{fig-s1} represents the schematic diagram of NiFe$_2$O$_4$ molecular device.

\subsection{Band structure of NiFe$_2$O$_4$ nanowire}
\begin{figure}[!h]
\centerline{\includegraphics[width=0.65\textwidth]{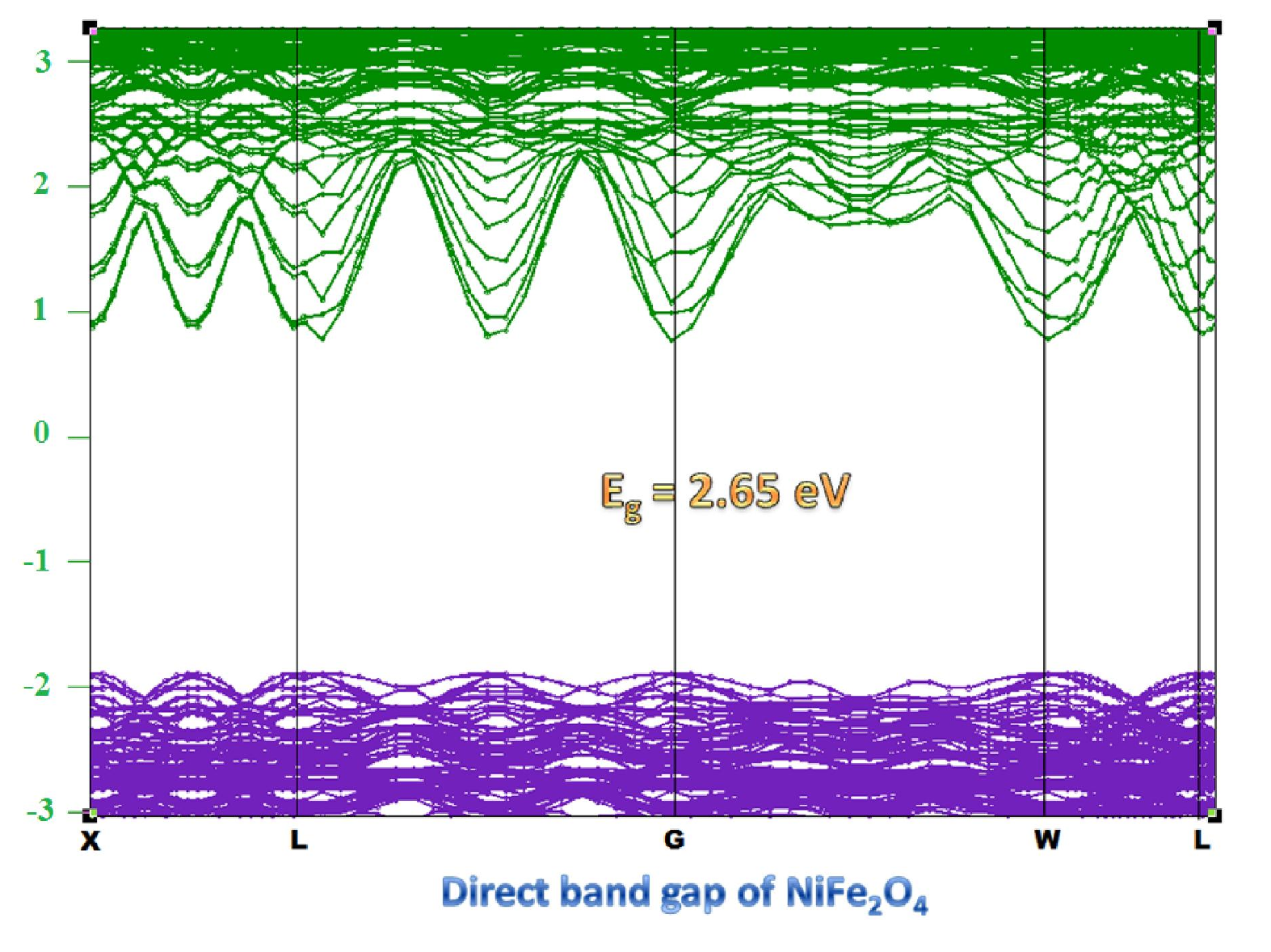}}
\caption{(Color online) Band structure of NiFe$_2$O$_4$ nanowire.} \label{fig-s2}
\end{figure}

The band structure of NiFe$_2$O$_4$ nanowire provides the insights on the materials properties of NiFe$_2$O$_4$ nanowire. The band structure of NiFe$_2$O$_4$ nanowire can be described in terms of conducting channels across the Fermi energy level ($E_{\text F}$) between the conduction band and the valence band \cite{48}. Figure~\ref{fig-s2} represents the band structure of NiFe$_2$O$_4$ nanowire. From the observation, it is known that NiFe$_2$O$_4$ nanowire has the band gap of 2.65~eV for the whole nanostructure, which exactly matches with the reported theoretical work \cite{49}. The experimental direct band gap value of NiFe$_2$O$_4$ is 2.5~eV, which is almost equal to the obtained theoretical value as shown in figure~\ref{fig-s2}. Thus, it can be suggested that SIESTA may be used as a significant computational tool for studying electronic properties of nanostructured materials with suitable basis sets. Moreover, the band gap of 2.65~eV for NiFe$_2$O$_4$ is one of the favorable conditions for the application in electronic devices.

\subsection{Density of states and electron density across NiFe$_2$O$_4$ nanowire device}
\begin{figure}[!t]
\vspace{-3mm}
\centerline{\includegraphics[width=0.7\textwidth]{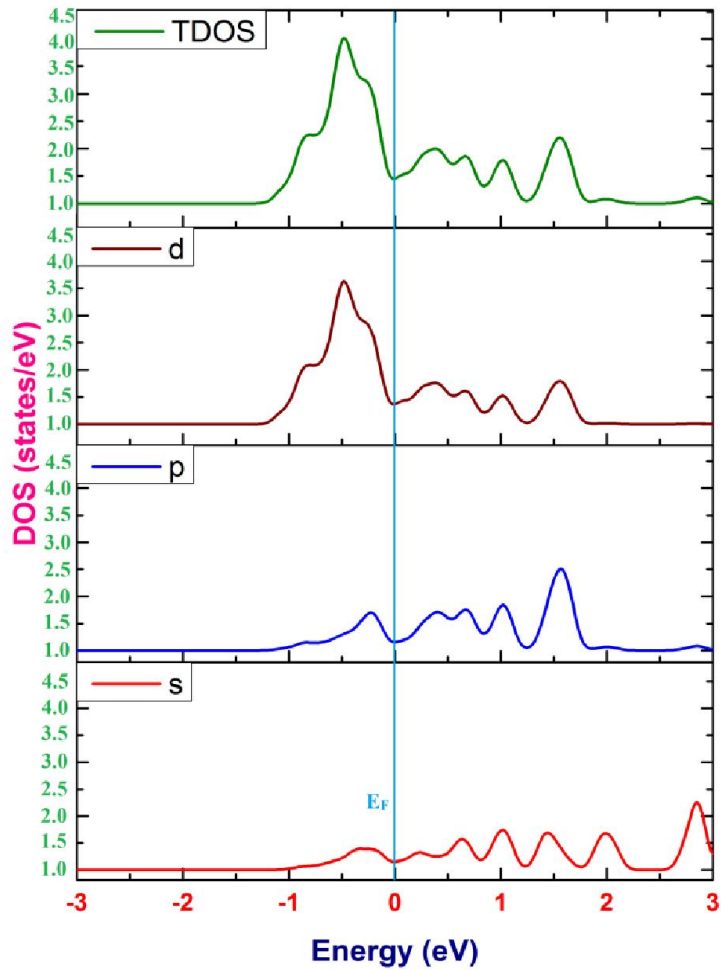}}
\vspace{-4mm}
\caption{(Color online) PDOS spectrum of NiFe$_2$O$_4$.} \label{fig-s3}
\end{figure}

The density of states (DOS) spectrum provides a clear picture regarding the density of charge in energy intervals along NiFe$_2$O$_4$ nanowire \cite{50,51,52}. Besides, the variation in bias voltage along NiFe$_2$O$_4$ nanowire leads to the change of the density of charge in the energy interval. In the present work, the variation in DOS is observed only beyond a threshold voltage of 2.5~V; which yields a significant change in the density of charge. On behalf of this reason, the bias voltage from 2.5~V to 7.5~V is carried out in the present study. In addition, the Fermi level ($E_{\text F}$) is kept at zero, since the bias window between right-hand and left-hand electrode is set as $-V/2$, $V/2$ in NiFe$_2$O$_4$ nanowire device. Figure~\ref{fig-s3} illustrates the projected density of states (PDOS) of NiFe$_2$O$_4$ base material. The base material refers to the basic element for building the molecular device. In the present work, NiFe$_2$O$_4$ is the base material that is used as electrodes and scattering region in the molecular device.
 Moreover, the major contribution in PDOS spectrum arises from $d$ orbitals of Ni and Fe, whereas for O, it is due to $p$ orbitals as observed in total DOS. The peak maxima at different energy levels are governed by the orbital overlapping of $d$ and $p$ orbital projected in NiFe$_2$O$_4$ base material. Furthermore, the peak maxima are observed near the Fermi level, which upon applying the bias voltage results in the transition of electrons from the valence band to the conduction band.

\begin{figure}[!t]
\centerline{\includegraphics[width=0.65\textwidth]{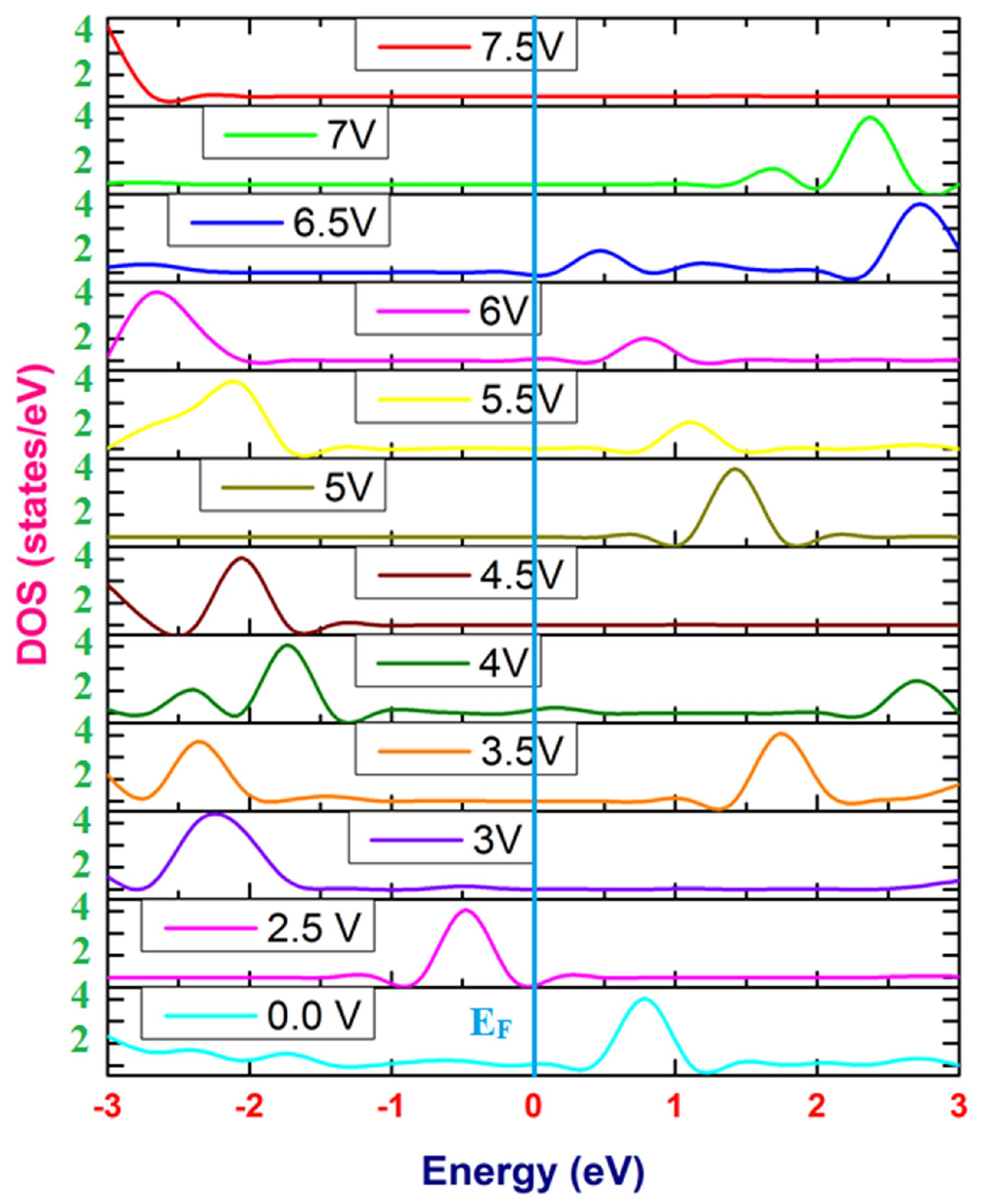}}
\caption{(Color online) Device DOS of NiFe$_2$O$_4$ nanowire. } \label{fig-s4}
\end{figure}

Figure~\ref{fig-s4} refers the device density of states spectrum for 0.0~V, 2.5~V, 3.0~V, 3.5~V, 4.0~V, 4.5~V, 5.0~V, 5.5~V, 6.0~V, 6.5~V, 7.0~V and 7.5~V bias. For 0~V bias, the DOS spectrum across NiFe$_2$O$_4$ nanowire is observed to be more in the conduction band than in the valence band. The peak maximum is recorded to be around 0.85~eV in the conduction band. Interestingly, at zero bias voltage condition, the peaks arise due to the mismatch of electronic chemical potential between the electrodes, thus localization of charges is observed in the conduction band.

\begin{figure}[!t]
\vspace{-4mm}
\centerline{\includegraphics[width=1.05\textwidth]{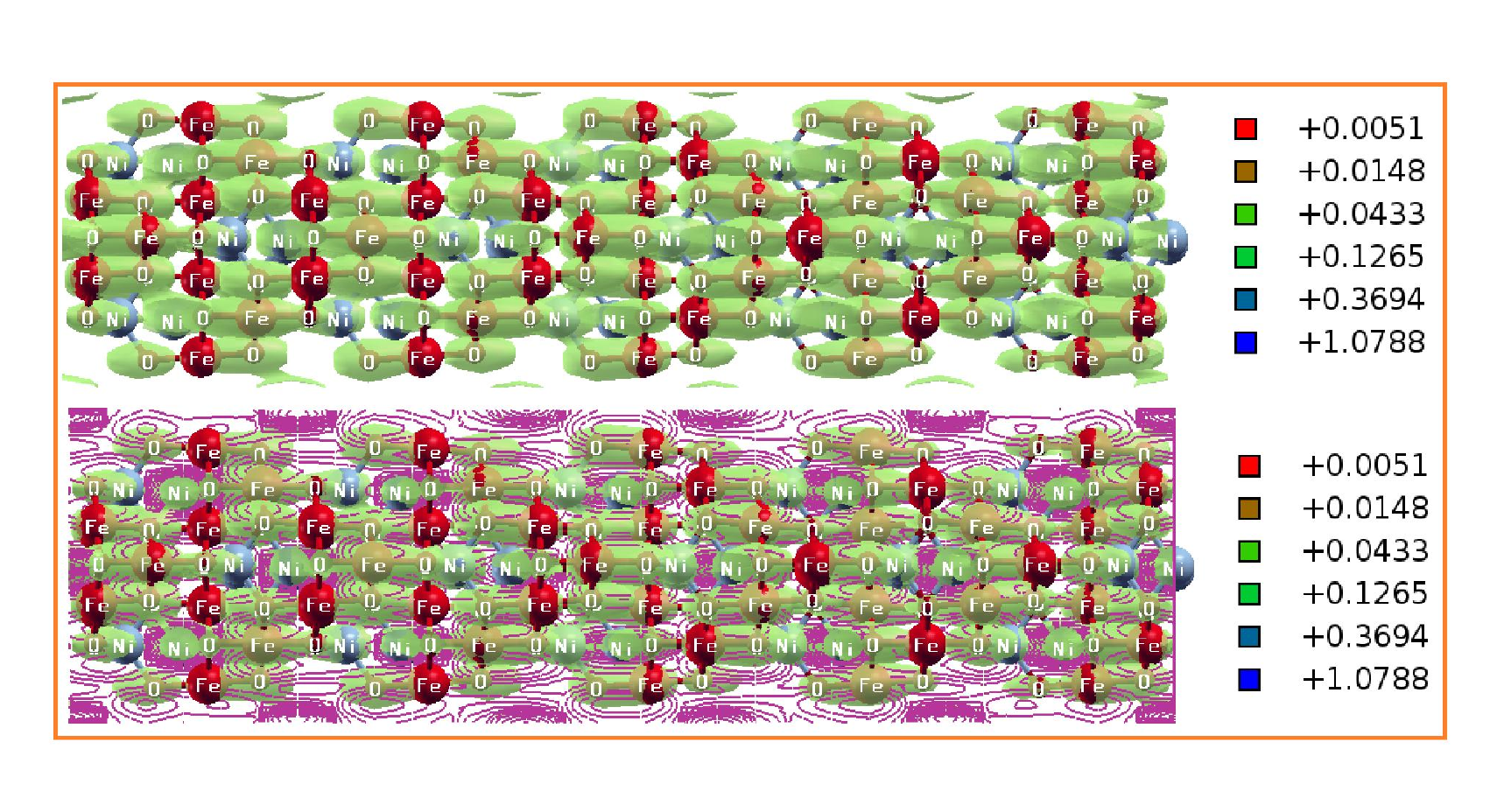}}
\vspace{-4mm}
\caption{(Color online) Electron density of NiFe$_2$O$_4$ nanowire. } \label{fig-s5}
\end{figure}

There is no significant peak maximum observed in the valence band of NiFe$_2$O$_4$ nanowire device at 0~V. Furthermore, on applying the bias voltage of 2.5~V across the electrodes, the localization of charges is recorded near the Fermi level as shown in figure~\ref{fig-s3}. In addition, increasing the bias voltage to 3.0~V across NiFe$_2$O$_4$ nanowire device, results in peak maximum at $-$2.5~eV in the valence band. When the bias voltage is set to 3.5~V, localization of charges is observed on both the valence band and the conduction band within the energy interval of $-$2.4 and 1.75~eV, respectively. This infers that the bias voltage drives the charges from the valence band to the conduction band along NiFe$_2$O$_4$ scattering region. The same trend is observed at the bias voltage of 4.0~V. The only difference is that the localization of charges is shifted towards the conduction band on increasing the bias voltage. When the bias voltage is switched to 4.5~V, the localization of charges is noticed in the valence band at $-$2.1~eV. However, the charge transition takes place for the bias voltage of 5.0~V and the peak is observed at 1.4~eV. In the case of bias voltage for 5.5 and 6.0~V, the peak maxima are observed on both the conduction band and the valence band. By contrast, the localization of charges is observed only on the conduction band at different energy intervals in the case of 6.5 and 7.0~V bias voltages. Thus, it is inferred that the density of charge along NiFe$_2$O$_4$ nanowire device can be finely tuned with the bias voltage.
The electron density across NiFe$_2$O$_4$ nanowire is shown in figure~\ref{fig-s5}. The density of electrons is observed to be more in oxygen sites than in iron and nickel sites along NiFe$_2$O$_4$ nanostructure. Since the atomic number of the oxygen atom is eight and it is belongs to the group VIA element, due to the electronegative property of oxygen, it results in the accumulation of more electrons across oxygen sites in NiFe$_2$O$_4$ nanowire. One of the most significant chemical properties of the oxygen atom is the electronegativity property, which is accredited as the tendency of oxygen to attract electrons towards it. Moreover, the electron density is larger along the oxygen sites owing to the electronic configuration of the oxygen atom when bonding with nickel and iron atoms in NiFe$_2$O$_4$ nanowire. Besides, the electronegativity of the oxygen atom is also influenced by the distance between nucleus and valence electrons in NiFe$_2$O$_4$ nanowire. The electron density provides the insight on the chemical and electronic properties of NiFe$_2$O$_4$ nanowire.

\subsection{Transport properties of NiFe$_2$O$_4$ nanowire device}
The electronic transport of NiFe$_2$O$_4$ molecular device can be ascribed in terms of transmission spectrum \cite{53,54,55}. The transport characteristics of NiFe$_2$O$_4$ nanowire devices are investigated using TranSIESTA module in SIESTA package. The transmission function $T(E,V)$ of NiFe$_2$O$_4$ molecular device can be expressed as the sum of the probabilities of transmission for all the channels at energy $E$ beneath external bias voltage $V$ as shown in equation (\ref{1})
\begin{equation}
	T (E,V) = \Tr \big[ \Gamma_{\text L} (V) G^{\text R} (E,V) \Gamma_{\text R}(V)G^{\text A}(E,V)\big],
\label{1}
\end{equation}
where $\Gamma_{\text{R,L}}$ is the coupling function of the right-hand and left-hand self-energies, respectively. $G^{\text A}$ and $G^{\text R}$ are the advanced and retarded Green's function. Furthermore, the molecular orbitals nearer to the Fermi energy level ($E_{\text F}$) facilitate the electronic transport across NiFe$_2$O$_4$ nanowire even for the low bias voltage. The general relation between the conductance and transmission probability under zero bias condition is given as
\begin{equation}
G = G_0T(E,V=0),
\end{equation}
where $G_0$ is the quantum unit of conductance and it is equal to $2e^2/h$, $h$ is Planck's constant and $e$ is the electronic charge. The potential of $-V/2$ and $+V/2$ is maintained between the right-hand and left-hand electrode across NiFe$_2$O$_4$ molecular device, respectively.

The current through the NiFe$_2$O$_4$ nanowire device can be calculated from the Landauer-B\"{u}ttiker formula \cite{56}
\begin{equation}
I(V)= \frac{2e^2}{h} \int_{\mu_{\text L}}^{\mu_{\text R}} T \left(E, V_{\text b}\right)\rd E,	
\end{equation}
where $e$ is the elementary charge, $2e^2/h$ is the quantum conductance, $\mu_{\text{L,R}}$ is the electrochemical potential of left-hand and right-hand electrode, respectively.
	
\begin{figure}[!t]
\vspace{-5mm}
\centerline{\includegraphics[width=0.65\textwidth]{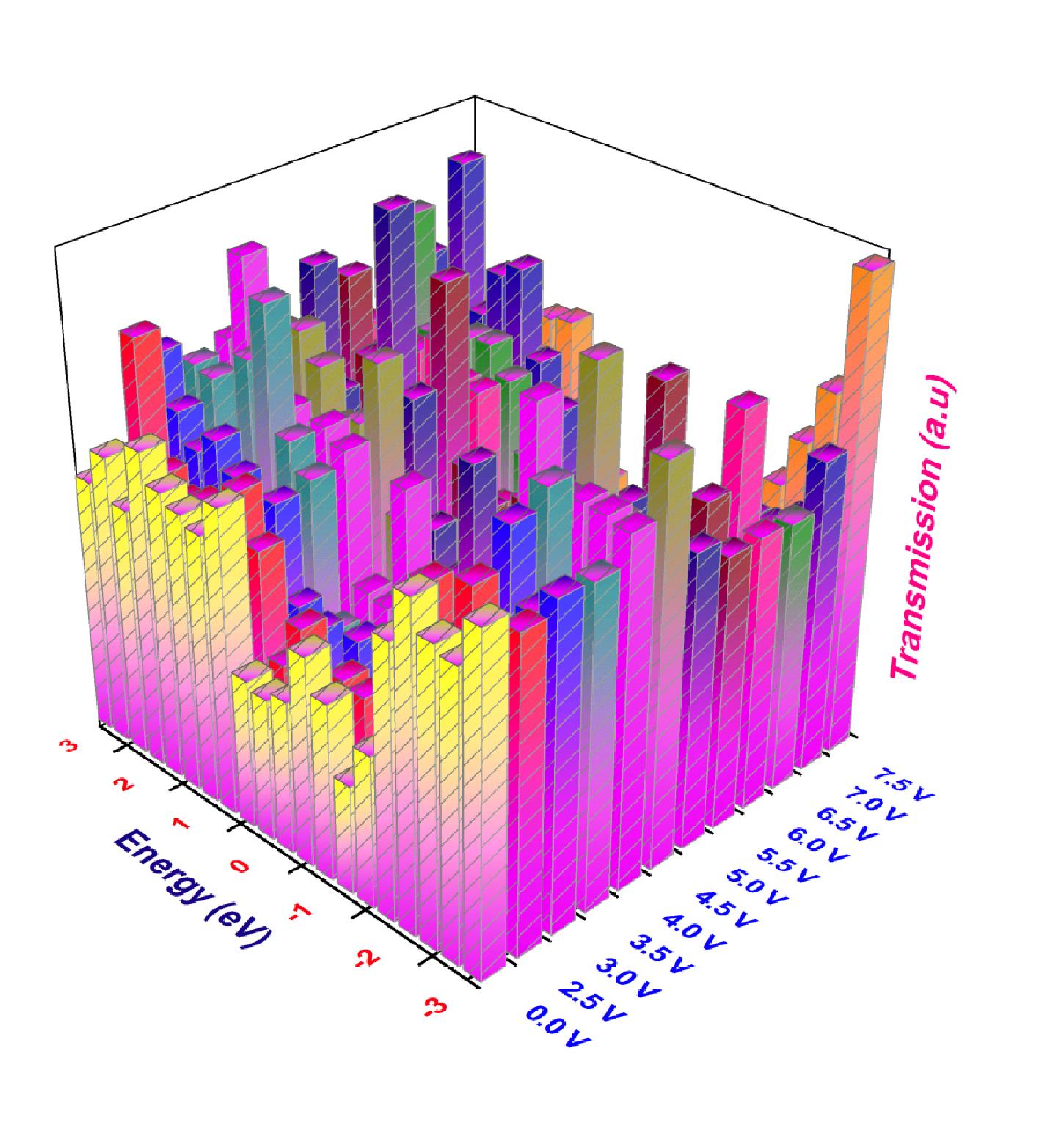}}
\vspace{-7mm}
\caption{(Color online) Transmission spectrum of NiFe$_2$O$_4$ nanowire. } \label{fig-s6}
\end{figure}

When zero bias is set across NiFe$_2$O$_4$ nanowire device, the Fermi level of left-hand electrode and right-hand electrode gets aligned and the electronic transmission between right-hand and left-hand electrode is equal in both directions, hence Fermi level is considered as zero. Figure~\ref{fig-s6} depicts the transmission spectrum of NiFe$_2$O$_4$ nanowire for different bias voltages. (The transmission spectrum is drawn in a three dimensional multi-curve fashion; the magnitude is taken into consideration along $y$ axis.) Besides, the transmission peaks recorded for the zero bias voltage are owing to the mismatch in the electronic chemical potential across right-hand electrode and left-hand electrode in NiFe$_2$O$_4$ nanowire device. By contrast, low peak amplitude is recorded in the conduction band.  On applying the bias voltage above zero, the molecular orbitals in NiFe$_2$O$_4$ nanowire get delocalized.  In that case, the mobility is recorded to be more in these energy intervals in the transmission spectrum \cite{57}. This gives rise to a certain peak maximum in the transmission spectrum of NiFe$_2$O$_4$ nanowire device \cite{58}. However, on increasing the bias voltage across NiFe$_2$O$_4$ scattering region, the transmission pathways increase along the NiFe$_2$O$_4$ nanowire; this gives rise to a shift in the peak maximum \cite{59}. Besides, when the bias voltage of 2.5~V is applied between the electrodes, the peak maximum is observed around 2.6~eV.

The increase of the bias voltage leads to the flow of electrons along the scattering region and the peak maximum moves towards the conduction band for the potential difference of 2.5~V. In the case of 3.0~V, the peak maximum is observed at $-$2.5~eV on the valence band and the peak gets shifted to the conduction band on applying the bias voltage of 3.5~V as shown in figure~\ref{fig-s5}. Furthermore, due to the transition of electrons across the scattering region along NiFe$_2$O$_4$, the peak maximum shifts to a different energy interval on varying the bias voltage. The applied bias voltage drives the electrons across the NiFe$_2$O$_4$ molecular device, in which the peak maximum gets shifted. For the applied bias of 4.0~V, the peak maximum is observed on both the valence band and the conduction band at $-$1.65 and 2.75~eV, respectively. Further increasing the bias voltage from 4.5 to 7.5~V, the peak maximum gets shifted along the valence band and the conduction band. The transmission spectrum has a peak maximum along different energy levels. The change in the current for different voltages should not be correlated directly with transmission spectrum with that of $I$--$V$ characteristics curve. The transmission spectrum indicates that the transmission of charges is larger for a particular energy interval to the applied bias voltage. However, the net current flowing through the molecular device depends on overall transmission for a different energy interval. This clearly suggests that the bias voltage is adequate enough for the transition of electrons along NiFe$_2$O$_4$ nanowire device and the transmission is governed by the applied bias voltage. Thus, it can be concluded that the transport property of NiFe$_2$O$_4$ nanowire device can be finely tuned by applying the proper bias voltage and can be used as a chemical sensor in microwave devices.

\subsection{$I$--$V$ characteristics of NiFe$_2$O$_4$ nanowire device}
\begin{figure}[!t]
\centerline{\includegraphics[width=0.65\textwidth]{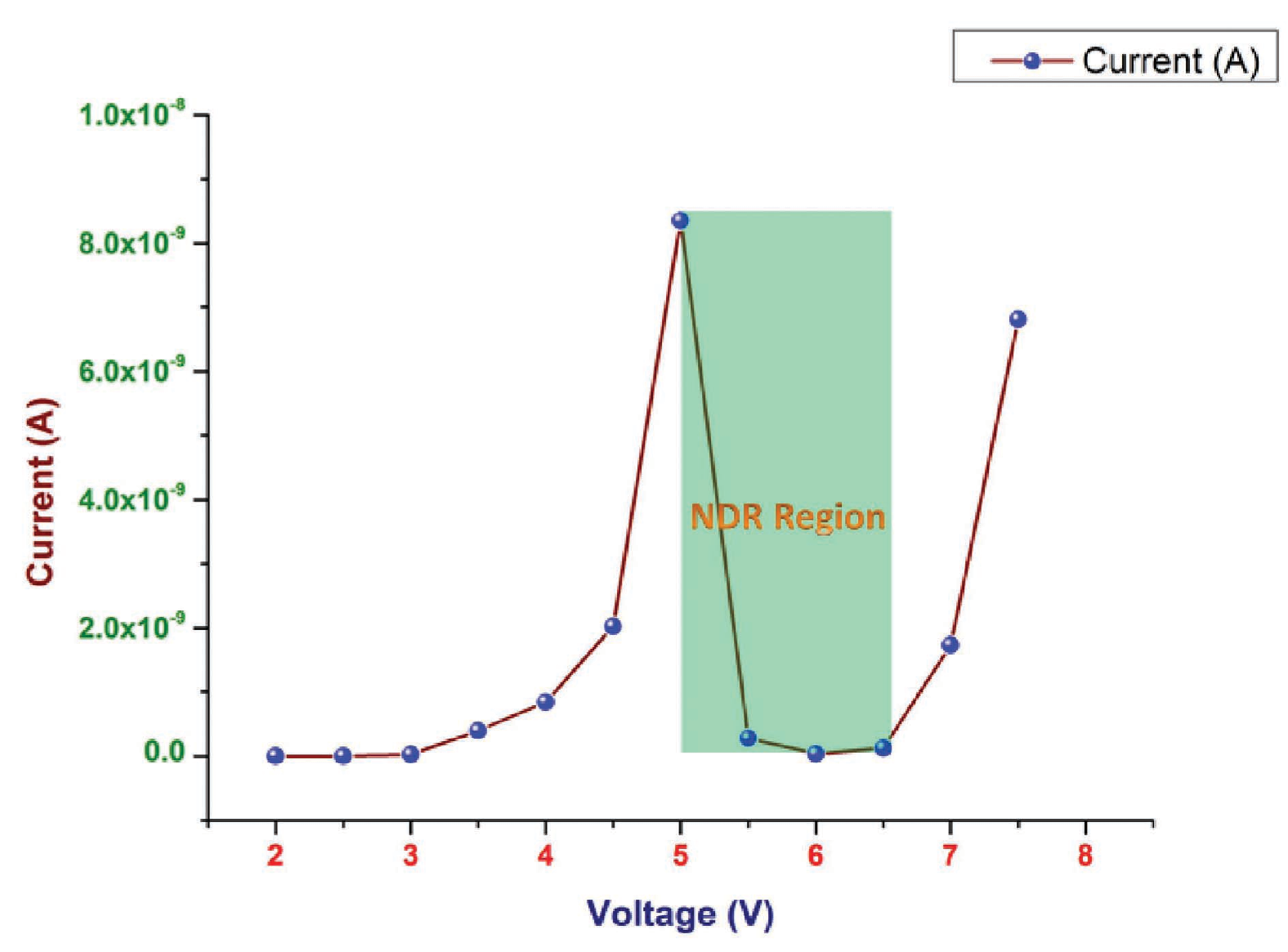}}
\caption{(Color online) $I$--$V$ Characteristics of NiFe$_2$O$_4$ nanowire. } \label{fig-s7}
\end{figure}

Negative differential resistance behaviour is the most significant electronic transport property for various electronic components \cite{26}. In the present study, the NDR behaviour is observed in the $I$--$V$ characteristics of NiFe$_2$O$_4$ nanowire as shown in figure~\ref{fig-s7}. The behaviour of NiFe$_2$O$_4$ molecular device is similar to that of an n-type semiconductor. At the beginning, the current flowing through NiFe$_2$O$_4$ nanowire device shows almost a linear increase across NiFe$_2$O$_4$ scattering region on increasing the bias voltage. Up to the threshold limit of 5~V bias, the current increases linearly for the applied bias. The NDR is observed for the bias voltage of 5.0~V to 6.0~V. Moreover, when NiFe$_2$O$_4$ nanowire device is operated in this bias voltage, it exhibits NDR. Further increasing the bias voltage beyond 6.0~V along NiFe$_2$O$_4$ nanowire device, the NDR behaviour vanishes and the device obeys the ohm’s law.

\begin{figure}[!b]
\vspace{-8mm}
\centerline{\includegraphics[width=0.95\textwidth]{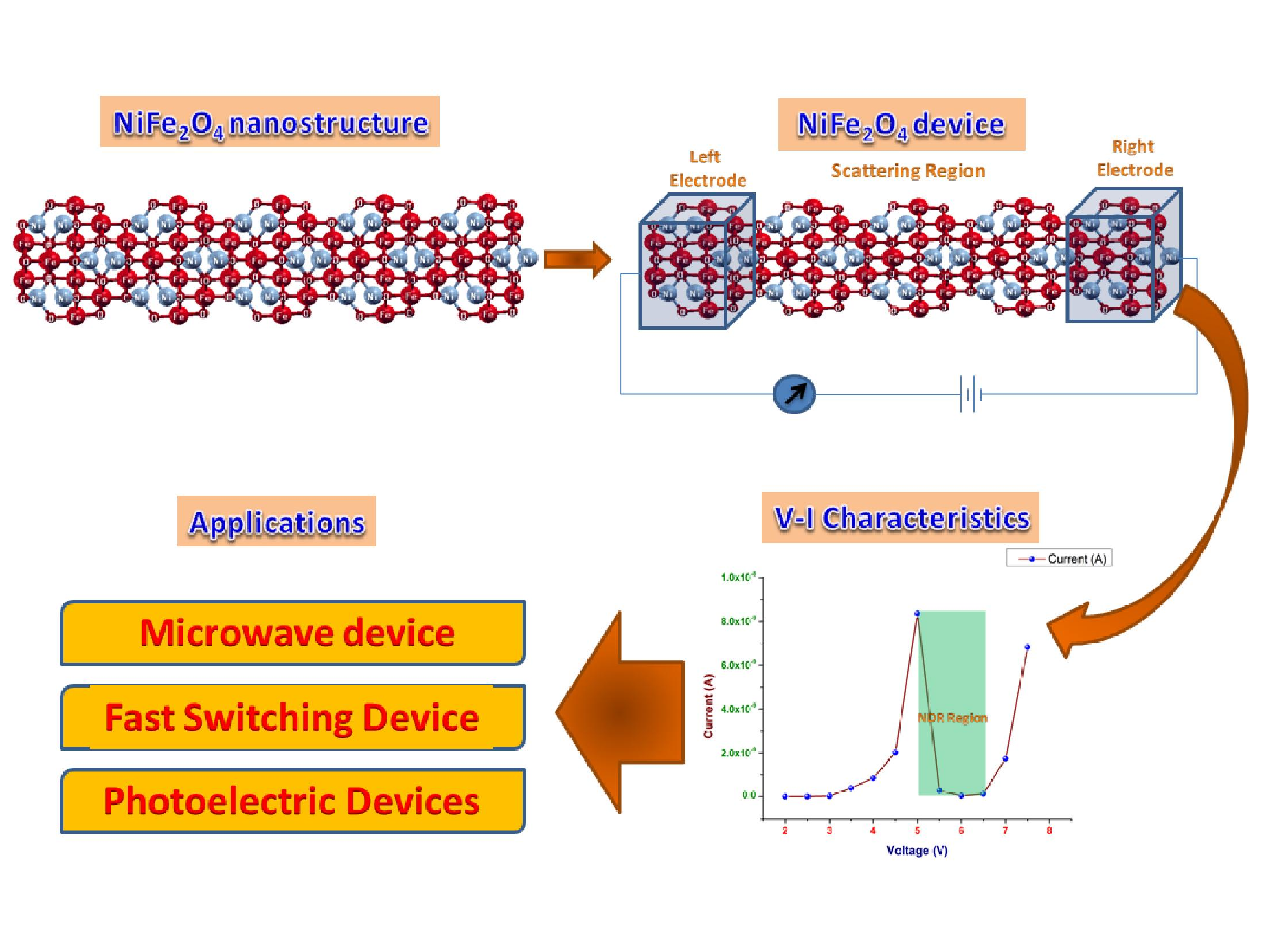}}
\vspace{-7mm}
\caption{(Color online) Schematic diagram of NiFe$_2$O$_4$ as NDR device.  } \label{fig-s8}
\end{figure}

In the present work, N-shaped NDR is observed for NiFe$_2$O$_4$ molecular device. The NDR behaviour in NiFe$_2$O$_4$ nanowire device originates from the inhibition of the conduction channels at a certain bias condition \cite{60}. Besides, the frontier orbitals localized in any part of the scattering region will not contribute to the transmission spectra and the current conduction may be suppressed.  By contrast, a completely delocalized molecular orbital may contribute more to the transmission probabilities than that of the localized one in NiFe$_2$O$_4$ nanowire device. Figure~\ref{fig-s8} illustrates the schematic diagram of NiFe$_2$O$_4$ nanowire device, which can be used as NDR device. Li et al. \cite{61} observed the N-shaped NDR in GaAs-based modulation-doped FET along with InAs quantum dots. Xu et al. \cite{62} reported a similar N-shaped negative differential resistance in GaAs-based modulation-doped FET with InAs quantum dots.
The NDR effect observed in the device is not only related to a single physical mechanism. Many phenomena give rise to the NDR property, namely tunneling, Coulomb blockade, Gunn effect \cite{63}, metal and semiconductor contact, charge storage and geometry of the nanodevice. Furthermore, the cylindrical geometry and high surface-to-volume ratio of nanowire results in deep penetration of the surface charge, which largely affect the conduction property of nanowire. From the Landauer-B\"uttiker relation, it is well known that the current through the device depends on $T(E,V)$. The current in the NiFe$_2$O$_4$ device is the integral of the transmission coefficient in the bias window of [$-V/2$, $V/2$]. In the present work, the NDR effect is observed in the bias voltage of around 5~V to 6~V. Moreover,  the device DOS (figure~\ref{fig-s4}) indicates a peak in the conduction band for 5~V at the energy level of 1.4~eV, whereas for 5.5~V and 6~V bias, the peaks are observed both in the conduction band and in the valence band. Thus, for the applied bias voltage of 5~V, the current increases drastically, and the further increase in the bias voltage gives rise to a decrease in the current due to the Coulomb blockade that arise due to the geometry of the device. Furthermore, for the bias of 5~V to 6~V, the bias window makes transition of electrons between the highest occupied molecular orbital (HOMO) and the lowest unoccupied molecular orbital (LUMO) decrease. The decrease in the transmission (figure~\ref{fig-s6}) takes place because a larger wave function overlaps between the scattering region and electrodes, the degree of coupling between the molecular orbitals and electrodes becomes weaker with an increase in the bias voltage beyond 5~V. Moreover, such a decrease may not be compensated by the increase in the bias voltage, thus the integral area gets smaller. However, on further increasing the bias voltage beyond 6~V, the degree of coupling between the electrodes and scattering region is overcome by the bias voltage and the current increases further more for the applied bias voltage.

The negative differential resistance properties are observed on various materials with different morphology such as ZnO nanorod, porous silicon devices and graphene nanoribbon FET \cite{38, 64, 65}. The NDR property of NiFe$_2$O$_4$ nanowire device is similar to the reported works, which further strengthens the present work. Thus, the negative differential resistance property of NiFe$_2$O$_4$ nanowire can be finely tuned by applying a proper bias voltage.

\section{Conclusions}

In the present study, NiFe$_2$O$_4$ nanowire based molecular device is studied using DFT method. Under various bias voltages, the electronic transport properties of inverse spinel NiFe$_2$O$_4$ nanowire device is investigated. The density of charges among different energy intervals of NiFe$_2$O$_4$ nanowire is clearly studied with the help of projected density of states spectrum. Moreover, the peak maximum is observed on both the valence band and the conduction band, which is influenced by the applied bias voltage. The electron density is observed to be more on oxygen sites along NiFe$_2$O$_4$ nanowire. The transmission spectrum of NiFe$_2$O$_4$ nanowire device shows a larger peak maximum in the valence band at the zero bias condition. However, on increasing the bias voltage, a larger peak maximum in the conduction band is observed, which clearly suggests that the bias voltage drives the charges towards the conduction band. The NDR properties of NiFe$_2$O$_4$ nanowire are investigated using $I$--$V$ characteristics. The NDR property of NiFe$_2$O$_4$ nanowire device depends on the applied bias voltage. Thus, the NDR property can be finely tuned with the bias voltage.  The findings of the present work in NiFe$_2$O$_4$ nanowire device can be used as NDR device, which may find its potential application in microwave devices, memory devices and in fast switching devices.

\ukrainianpart

\title{Дослідження з перших принципів нікель-феритового нанодротового пристрою з негативним диференційним опором}%
\author{В. Нагараджан, Р. Чандірамулі}
\address{Школа електротехніки та електроніки, Академія мистецтв, наукових і технологічних досліджень  Шанмуга (університет SASTRA), Танджавур, Таміл-Наду --- 613 401, Індія}

\makeukrtitle

\begin{abstract}
Електронні властивості  NiFe$_2$O$_4$ нанодротового пристрою  досліджується з використанням методу нерівноважних функцій Гріна в комбінації з теорією функціоналу густини. Властивості електронного переносу  NiFe$_2$O$_4$ нанодроту вивчаються в термінах густини станів, спектру трансмісії та  $I$--$V$ характеристик. Густина станів змінюється при прикладанні зміщувальної напруги через NiFe$_2$O$_4$ нанодротовий пристрій, густина заряду спостерігається як у валентній зоні, так і в зоні провідності при збільшенні напруги зміщення. Спектр трансмісії  NiFe$_2$O$_4$ нанодротового пристрою дає уявлення про перехід електронів на різних енергетичних інтервалах. Результати даної роботи наводять на думку, що  NiFe$_2$O$_4$ нанодротовий пристрій може бути використаний як негативний диференційний опір, і ця його властивість може бути регульована  за допомогою напруги зміщення, що може мати потенційне використання у мікрохвильових пристроях, пристроях пам'яті і в перемикальних пристроях.
\keywords нікель ферит, нанодріт, негативний диференційний опір, густина станів, електронна густина %
\end{abstract}

\end{document}